\documentclass[journal=apchd5,manuscript=article]{achemso}

\usepackage[version=3]{mhchem} 
\usepackage[T1]{fontenc}       
\usepackage{bm}

\usepackage{ulem}
\usepackage{siunitx}

\usepackage{color}

\title{Heuristic Modeling of Strong Coupling in Plasmonic Resonators}

\author{G\"unter Kewes}
\affiliation{AG Nanooptik, Humboldt-Universit\"at zu Berlin, 12846 Berlin, Germany}
\email{gkewes@physik.hu-berlin.de}
\author{Felix Binkowski}
\affiliation{Zuse Institute Berlin, 14195 Berlin, Germany}
\author{Sven Burger}
\affiliation{Zuse Institute Berlin, 14195 Berlin, Germany}
\alsoaffiliation{JCMwave GmbH, 14050 Berlin, Germany}
\author{Lin Zschiedrich}
\affiliation{JCMwave GmbH, 14050 Berlin, Germany}
\author{Oliver Benson}
\affiliation{AG Nanooptik, Humboldt-Universit\"at zu Berlin, 12846 Berlin, Germany}

\date{\today}
\begin{document}


\begin{abstract}
Strong coupling of plasmonic excitations and dipolar emitters, such as organic molecules, have been studied extensively in the last years. The questions whether strong coupling can be achieved with a single molecule only and how this is unambiguously proven are still under debate. 
A critical issue of plasmonic in contrast to photonic systems is additional excitonic line broadening, which is often neglected when modeling such systems.
This has led to too optimistic design predictions or incorrect interpretation of ambiguous experimental data, for example in models relying on Maxwell solvers without self-consistent incorporation of line broadening effects. 
In this paper, we present a heuristic modeling approach for strongly coupled systems based on plasmonic nanoparticles and dipolar emitters that accounts for such broadening and elucidates on recent experiments with single emitters.
We explicitly focus on a clear and intuitive classical description that utilizes well-established methods, easy to use within typical Maxwell solvers. The heuristic model (i) provides experimentally relevant numbers like emitter densities and spectra (ii) allows to discriminate systems, which can reach the strong coupling regime from those, which can not (iii) allows to identify optimization routes and (iv) nicely matches with experimental findings. 
In particular, we employ an approach related to quasi normal modes and extinction simulations where the excitonic system is represented by a frequency dependent permittivity. As examples, we investigate 
two configurations with many, but also single emitters, which have been studied in recent experiments.  
\end{abstract}


\maketitle


Strong coupling (SC) of single emitters and photonic cavities \cite{Vahala2003} has been theoretically studied and experimentally demonstrated in various configurations in the last decade \cite{Barnes2018SpecialCavities}. More recently, experimental work on strongly coupled systems has stimulated pronounced activity also in the plasmonics community \cite{Torma2015StrongReview}. Interestingly, SC can be reached with relatively low experimental effort and at ambient conditions with propagating surface plasmon polaritons (SPPs). This has triggered speculation whether SC can be pushed towards real world applications. SC could tailor optical responses, reduce the energy consumption of endothermic chemical reactions or it might create exciton-plasmon-polariton condensates \cite{Plumhof2014Room-temperaturePolymer,Hutchison2012ModifyingFields,Shalabney2015CoherentMode}.

Researchers in the field of nanooptics and plasmonics now consider replacing photonic cavities by fully nanoscopic systems, i.e., plasmonic nanoresonators \cite{Torma2015StrongReview,Baranov2018NovelInteractions}. These support localized plasmon polaritons (we denote them here as resonant plasmonic modes) and promise a significant reduction of the spatial footprint of strongly coupled systems due to their extremely small (mode) volumes. The low quality ($Q$) factors  of plasmonic modes are no principle restriction since they are counterbalanced by much stronger interaction strength due to larger field confinement. For applications in ultrafast physics, a lower $Q$ and thus larger bandwidth even renders advantageous as compared to narrow-band high-$Q$ dielectric systems. Additionally, the broad bandwidth of plasmonic systems seems much better suited for SC at ambient conditions where emitters are broad-band, too. 

SC is characterized by a transition from an incoherent to a coherent emitter-field dynamics. New hybrid states form, indicated by a splitting of a previously degenerate peak or dip in scattering, absorption and photoluminescence (PL) spectra, respectively. This is, however, much harder to reveal in plasmonic systems due to their intrinsically broad features. A special challenge is the level of single plasmonic particles acting as plasmonic resonators coupled to only few emitters. In this case, dark-field scattering spectroscopy is often utilized while measuring the absorption and PL \cite{Wersall2017ObservationExcitons,Melnikau2016RabiJ-Aggregates} remains very challenging in contrast to ensemble measurements \cite{Melnikau2016RabiJ-Aggregates,Stete2017SignaturesEnvironment}. 

Plasmonic resonators based on nanoparticles are often regarded as nanoantennas, i.e., light concentrators or extractors. A prominent antenna effect of plasmonic particles is the enhancement of the absorption cross section of nearby dipole emitter (DE) by orders of magnitude compared to their reference value \cite{Bharadwaj2009OpticalAntennas}. The latter is typically measured independently, e.g., with an ensemble of DEs in solution. The enhanced absorption cross sections can indeed become so large that it competes with the scattering cross section of the antenna and "eats up" originally scattered light leaving a dip in the spectral response. In this case, even a single DE can represent a significant damping mechanism to the plasmonic mode yielding an overall increase of its linewidth. Such a broadening and the appearance of a central dip of the plasmonic modes scattering resonance are merely an antenna effect and must not be misinterpreted as SC. Actually, such effects were observed already some years ago and were called "plasmon quenching dips" \cite{Liu2007QuantizedTransfer}. As long as alternative interpretation of dips in scattering spectra can not be excluded, scattering experiments are \textit{a priori} inconclusive. 

A proof of SC would be the direct detection of the coherent energy exchange (Rabi oscillation) between the DE and the plasmonic resonator \cite{Vasa2013Real-timeJ-aggregates} by pump-probe experiments. However, due to the very fast time scale of Rabi oscillations in plasmonic systems, it is far from trivial to exclude other processes, like thermal, acoustic or non-linear photophysical effects, which may cause similar signatures.

Complementary to the experimental difficulties, which may easily lead to inconclusive results, there are obstacles with theoretical modeling. Sophisticated models for SC with plasmonic nanoresonators were introduced, but their predictions are hard to adapt to variations in a realistic multi-parameter experimental scenario. Remarkably, certain advanced theoretical methodologies are also exclusively carried out by individual groups making a critical evaluation difficult. Further, some theoretical approaches, e.g., full quantum treatments, are restricted to rather artificial situations with only individual DEs or with multiple, but indistinguishable DEs all experiencing the same coupling strength \cite{Richter2015NumericallyEmission,Gonzalez-Ballestero2016UncoupledProperties,Delga2014QuantumQuenching}. 

From an experimental point of view there is a lack of realistic easy-to-use models. Hence, to support their experimental findings, experimentalists may be inclined to use relatively simple simulation tools, which lack a self-consistent description of important effects yielding oversimplified estimates. There is a risk that such estimates are then compared to intrinsically inconclusive scattering spectra and taken as a proof for SC. We find that even effects that are well-known from similar plasmonic systems are frequently disregarded. The goal of this paper is to suggest heuristic and easy-to-use modeling methods which at the same time avoid over-simplifications. The strength of this model lies in its simplicity, transparency, ability to interpret experimental findings (discriminating weak from strong coupling) and to predict optimization routes including exact parameters for the materials.

\section*{Line Broadening in Strongly Coupled Systems}
\label{quenching}
Before we propose heuristic models for SC and apply them to various relevant situations in the next sections, we first discuss the problem of line broadening. Strong and weak coupling regimes in light-matter interaction are discriminated by the relative strength of the emitter-field coupling constant $g$, the emitter's decay rate (emission into non-cavity modes and into any other decay channel) $\gamma$, and the loss rate of the cavity or plasmon resonator  $\kappa$, which is half the linewidth of the cavity's or plasmon's resonance, respectively. In literature, there are several definitions for the SC regime. Most of them agree that the energy splitting measured as spectral distance of the new hybrid states formed by the localized field (photon or plasmon) and the DE should be larger than the linewidth of both uncoupled constituents: $\Omega\,\geq(\gamma$\,,$2\kappa$), where $\Omega=2g$ is the Rabi frequency. Frequently used explicit definitions are, e.g., $\Omega^2 > \frac{(2\kappa)^2}{2}+\frac{\gamma^2}{2}$ \cite{Torma2015StrongReview} and $\Omega^2 > 2\gamma\kappa$ \cite{Kimble1998StrongQED}; both conditions yield very similar results for the examples considered here (we use the latter one throughout this work).

A key difference of SC with plasmonic resonators as compared to optical cavities is that the rates of the individual constituents, resonator and DE, $\kappa$ and in particular $\gamma$, can significantly differ in the coupled and uncoupled case.  
For example, an atom brought into a milimeter-sized Fabry-P\'{e}rot cavity still emits spontaneously in free-space with its "natural" emission rate. However, for a DE implemented into a plasmonic mode's hotspot, new decay channels open up due to near-field coupling to other modes and in proximity to metal surfaces. While such additional line broadening has been discussed with respect to nanoantennas \cite{Moroz2010,Ruppin1982,Chew1987,Kim1988,Anger2006} and surface plasmon lasers (spasers) \cite{Kewes2017LimitationsSpasers}, it is often neglected or underestimated in the context of SC. Linewidths of DEs measured independently outside of or far away from plasmonic nanoresonators would lead to wrong or at least far too optimistic results when used as fixed input parameters in a numerical simulation. Another critical point is the calculation of the coupling constant $g$. Its magnitude depends on the mode volume $V_{\rm{M}}$ of the specific mode under consideration. To derive this quantity, it is necessary to normalize the electromagnetic field correctly. Although this theoretical problem can be regarded as completely solved now \cite{Sauvan2013,Kristensen2014,Lalanne2018LightResonances,Zschiedrich2018RieszNanoresonators}, one still finds inconsistent definitions of $V_{\rm{M}}$ in literature.

\begin{figure*}[ht]
\centering
\includegraphics[width=0.99\textwidth]{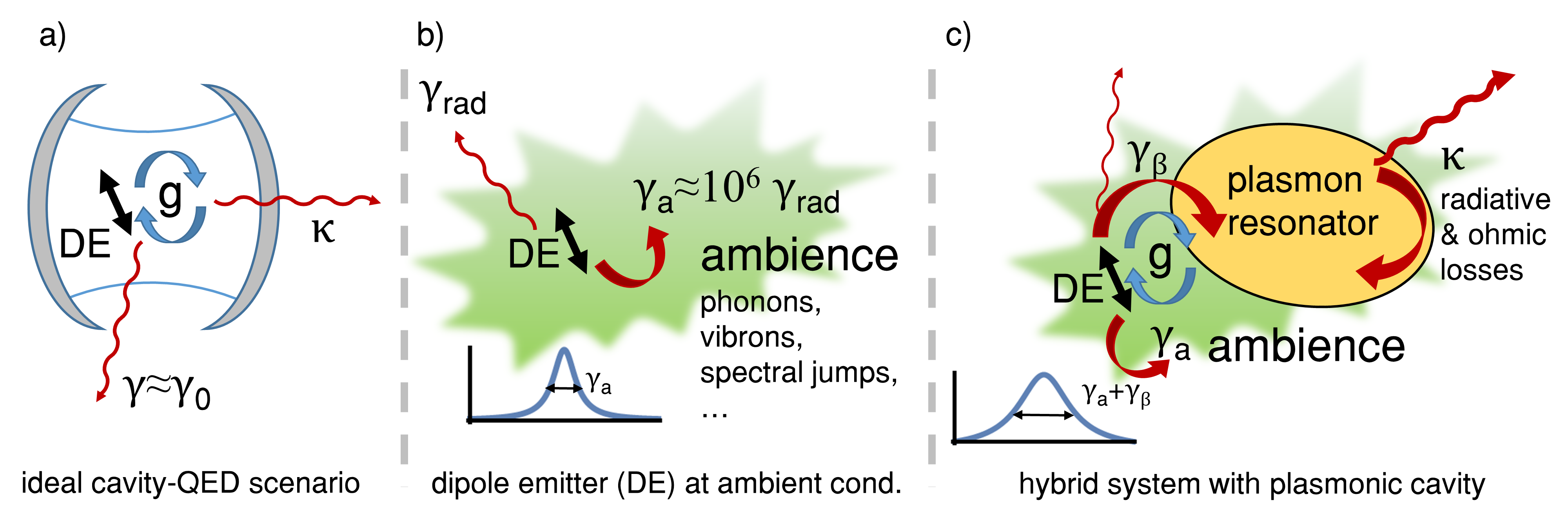}
\caption{\label{fig_physics_scheme} Schematics of interaction of dipole emitters (DEs) with different environments. (a) "Conventional" SC condition: a DE (typically an atom) inside a dielectric optical cavity. Coherent interaction is described via the emitter-field coupling constant $g$, whereas incoherent processes are determined by the photon decay rate $\kappa$ and the spontaneous emission rate $\gamma_{\rm{0}}$. For isolated atoms, $\gamma_{\rm{0}}$ is the only broadening mechanism of the DE, i.e., $\gamma=\gamma_{\rm{0}}$. In the strong coupling regime, $2g \geq (\kappa,\gamma)$ holds. (b) DE in a solid-state or condensed-phase matrix as often used in plasmonics. In addition to radiative decay $\gamma_{\rm{rad}}$ there are several other decay channels and broadening mechanisms. Under ambient conditions the linewidth $\gamma_{\rm{a}}$ is much broader than the limit given by the radiative decay.
(c) DE coupled to the near-field of a plasmonic resonator. In addition to the coherent coupling $g$, there are incoherent rates. Compared to the situation in b) the non-radiative decay of a DE $\gamma$ can be significantly enhanced with respect to its value outside of the plasmonic hotspot due to additional loss channels described by the rate $\gamma_{\rm{\beta}}$. 
}
\end{figure*}

We summarize the different scenarios for SC in optical cavities and plasmon resonators, respectively, in Fig.~\ref{fig_physics_scheme}. In "conventional" quantum optics with isolated atoms in optical cavities (Fig.~\ref{fig_physics_scheme}a) the decay rates are exclusively radiative. Then $\gamma$ is the spontaneous emission rate of the atom in free space $\gamma_{\rm{0}}$. 

The situation changes dramatically with DEs in a solid-state or condensed-phase matrix and plasmonic resonators (Fig.~\ref{fig_physics_scheme}b) and c)) for two reasons. First, in addition to spontaneous radiative emission described by the rate $\gamma_{\rm{rad}}$, there is coupling to the phonon bath of the solid and to vibronic modes of the DE itself [Fig.~\ref{fig_physics_scheme}b)]. This leads to large homogeneous line broadening far beyond the ideal Fourier-limit. At ambient conditions the linewidth $\gamma_{\rm{a}}$ exceeds the Fourier limit by a factor of $\approx 10^6$ \cite{Tamarat2000TenSpectroscopy.}. Since different DEs in a matrix typically see a different environment inhomogenoeus broadening further contributes to the linewidth. Second, coupling of the emitter to the plasmon resonator is dominated by near-fields [Fig.~\ref{fig_physics_scheme}c)]. Non-radiative rates due to energy transfer to an absorbing material are significant and reach values of around $\gamma_{\rm{\beta}} \approx 10^6 \gamma_{\rm{rad}}$  for emitters in sub-nanometer distance to metals \cite{Moroz2010,Ruppin1982,Chew1987,Kim1988}, which is the typical distance of DEs to metals in compounds for SC. Thus, the non-radiative decay of a DE in a hotspot of a plasmonic resonator $\gamma$, can be significantly enhanced with respect to its value outside of the hotspot, from $\gamma=\gamma_{\rm{a}}$ to $\gamma=\gamma_{\rm{a}}+\gamma_{\rm{\beta}}$.

In order to reach the SC regime with given DEs and a dielectric cavity it is the best strategy to enhance coupling into the resonant mode by reducing the mode volume and placing the DEs in a field maximum. In case of a plasmonic resonator, however, this will typically bring the DEs in closer proximity to the metal surface and increase additional decay rates $\gamma_{\rm{\beta}}$ making it harder to achieve SC. For SC in plasmonic resonators, one has to increase coupling to the mode and avoid opening up additional decay channels {\it at the same time}. For this reason, the optimum design of the resonator is much more crucial in the plasmonic case. In the following, we will discuss this in terms of the Purcell and $\beta$ factor, respectively.
Throughout the analysis we assume a homogeneously broadened DE characterized by a linewidth given by the radiative rate $\gamma_{\rm{0}}$ and the non-radiative decay rate $\gamma_{\rm{a}}$ {\it far away} from the plasmonic resonator, respectively. We further assume that the coupling of the DE to phonons or vibrons is constant. 
To derive realistic values for $g$ and $\gamma$ for a DE {\it in a hotspot} of a resonant plasmonic mode, one has to discriminate the modified local density of states (LDOS) due to a specific mode of the resonator, denoted $\rho_{\rm{M}}$, from the overall LDOS, denoted $\rho_{\rm{tot}}$. Both densities can be normalized to the LDOS in an isotropic, homogeneous host dielectric $\rho_{\rm{0}}$. The normalized total LDOS can be computed with Maxwell solvers by recording the power $P$ emitted by a Hertzian dipole via $\rho_{\rm{tot}}/\rho_{\rm{0}}=P/P_{\rm{0}}$ where $P_{\rm{0}}$ is the power emitted in the homogeneous host \cite{Novotny2006}. This LDOS includes any contribution, due to plasmonic near-fields but also due to other non-radiative channels. The modal LDOS $\rho_{\rm{M}}$ (and with it the mode volume $V_{\rm{M}}$) can be computed within the framework of quasi normal modes (QNMs) \cite{Sauvan2013,Kristensen2014} gk{\cite{Lalanne2018LightResonances}} or Riesz projections \cite{Zschiedrich2018RieszNanoresonators}.
We now introduce the common definition of the "modal Purcell factor" $\Gamma_{\rm{M}}=\rho_{\rm{M}}/\rho_{\rm{0}}$ and accordingly the "total Purcell factor" $\Gamma_{\rm{tot}}=\rho_{\rm{tot}}/\rho_{\rm{0}}$ \cite{Koenderink2017Single-PhotonNanoantennas}. The ratio of $\rho_{\rm{M}}$/$\rho_{\rm{tot}}$ is known as the $\beta$ factor and characterizes the coupling efficiency of a DE to a specific mode. Consequently, $1-\beta$ characterizes the coupling efficiency to all other channels. These other channels correspond to the additional decay rate $\gamma_{\rm{\beta}}$ mentioned above caused by the proximity of the metal surface. For SC, it is thus essential to increase $\Gamma_{\rm{M}}$ and $\beta$ at the same time.
From $\Gamma_{\rm{M}}$ one can compute the mode volume $V_{\rm{M}}$ as introduced in Ref.~\cite{Sauvan2013}. $V_{\rm{M}}$ is in general a complex number, which depends on the actual position and orientation of the emitter. In case of a DE perfectly aligned with respect to the resonator's mode and in resonance, one finds: 
\begin{equation}
\Gamma_{\rm{M}}=\frac{3}{4\pi^2}\left(\frac{\lambda}{n}\right)^3\text{Re}\left(\frac{Q}{V_{\rm{M}}}\right)
\label{purcell}
\end{equation}
where $Q=\frac{2\pi c}{\lambda \kappa} $ is the resonator's quality factor.
From $V_{\rm{M}}$ the coupling constant $g$ is derived if the dipole moment $\mu$ of the emitter is known: 
\begin{equation}
g=\mu \sqrt{\frac{\pi\hbar c N}{\lambda \epsilon_{\rm{d}} \epsilon_{\rm{0}}  V_{\rm{M}}}}=g_{\rm{0}}\sqrt{N} . 
\label{g}
\end{equation}
$N$ is the number of emitters coupling to the resonator and  $\epsilon_{\rm{d}}$ the relative permittivity. 
The additional decay channel $\gamma_{\rm{\beta}}$ introduced by the plasmonic nanoresonator can now be related to the total Purcell factor and thus to the 
$\beta$ factor via:

\begin{equation}
\gamma_{\rm{\beta}} = (\Gamma_{\rm{tot}}-\Gamma_{\rm{M}} )\gamma_{\rm{0}}=\frac{1-\beta}{\beta} \Gamma_{\rm{M}} \gamma_{\rm{0}} . 
\label{gamma}
\end{equation}

Here, we used that $\beta=\rho_{\rm{M}}/\rho_{\rm{tot}}$ is equal to $\Gamma_{\rm{M}}/\Gamma_{\rm{tot}}$ in the weak coupling regime. Consequently, Eq.~\ref{gamma} is only valid in the weak coupling regime. Thus, here $\beta$ and $\gamma_{\rm{\beta}}$ are only used to predict where SC will be reached, rather than for the description of systems in the SC regime.
The additional decay rate $\gamma_{\rm{\beta}}$ adds up with the rate $\gamma_{\rm{a}}$ to the total lost emission rate of the DE $\gamma=\gamma_{\rm{\beta}}+\gamma_{\rm{a}}$. Please note, that in specific situations with strong energy dissipation, i.e., energy exchange between different resonator modes, Purcell factors of individual modes can become negative which leads to unphysical $\beta$ factors. However, we believe that in most situations relevant for strong coupling with plasmonic resonators, energy dissipation plays a minor role (see common examples discussed below). Further, it is mandatory to analyze the modes supported by the plasmonic resonator thoroughly in any case, which would reveal such situations. A qualitative indication for systems with strong energy dissipation is the occurrence of a strongly non-Lorentzian lineshapes of $\rho_{\rm{tot}}$; a quantitative measure is the ratio of Im($V_M$)/Re($V_M$). In such a case still an analytic model as constructed in Ref.~\cite{Lalanne2018LightResonances} may be employed.

\begin{figure}
\includegraphics *[width=0.99\columnwidth]{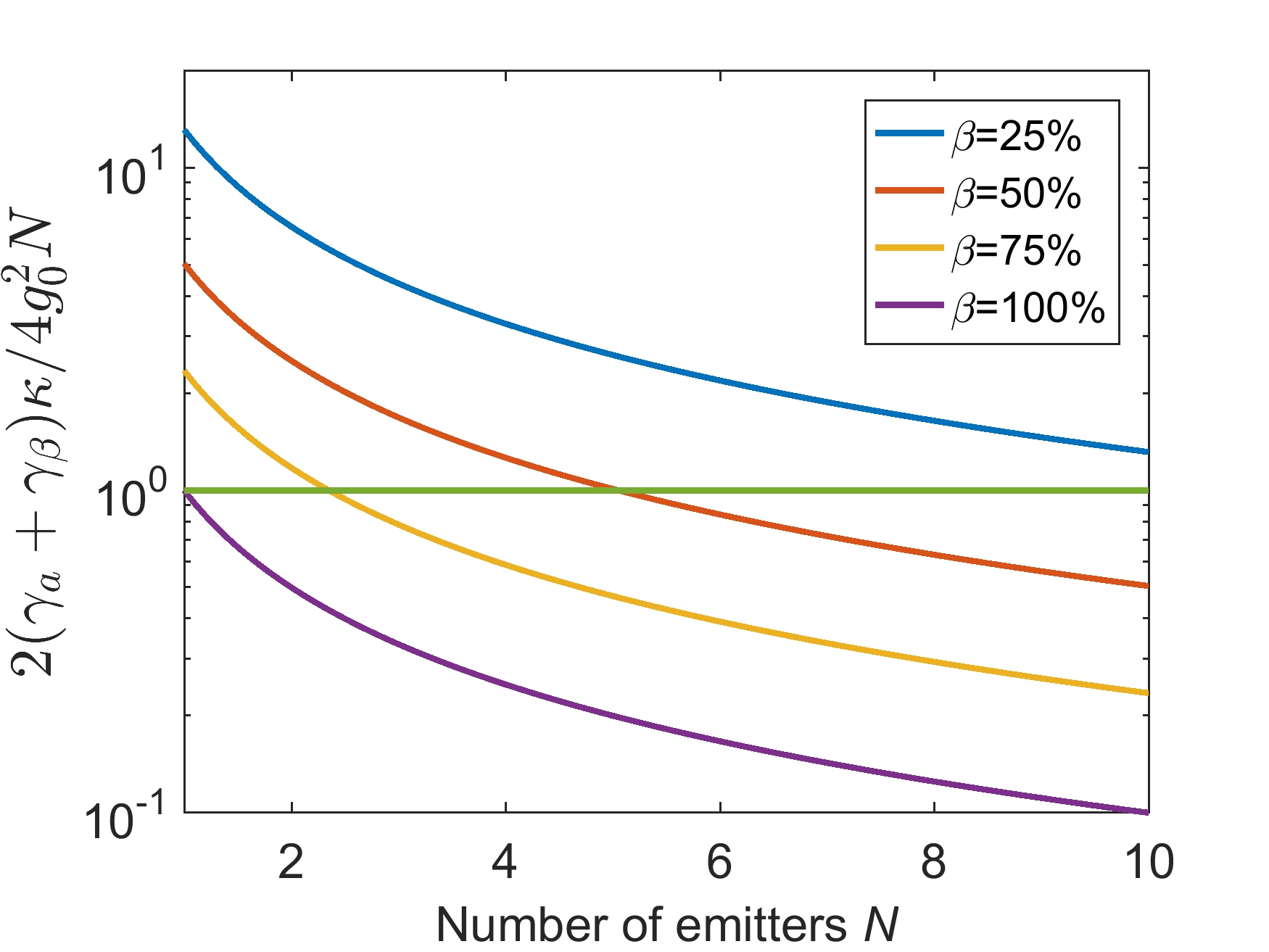}
\caption{\label{gamma_over_g_vs_N}
Dependence of the ratio of $2\gamma\kappa/(2g_0\sqrt{N})^2$ on the number of dipole emitters (DEs) $N$ for different $\beta$ factors. Values below unity (green horizontal line) indicate systems where strong coupling is reached.
}
\end{figure}

Starting with a specific plasmonic resonator (characterized by its resonance frequency $\omega_{\rm{res}}$, $\kappa$, and $V_{\rm{M}}$) and a specific DE (characterized by $\mu$ and $\gamma_{\rm{a}}$) one may derive the coupling constant per DE $g_{\rm{0}}$. 
In order to decrease the ratio of $2\gamma\kappa/(2g_0\sqrt{N})^2$ below unity, i.e., to reach the SC regime, the only way is to increase the number of DEs per mode volume, which is the density $N/V_{\rm{M}}$. Whenever $\beta$ is not reaching unity, the rise of $\gamma_{\rm{\beta}}$ will require more DEs to reach SC.
Figure~\ref{gamma_over_g_vs_N} shows $2\gamma\kappa/(2g_0\sqrt{N})^2$ versus the number $N$ of DEs for an example with realistic numbers. The virtual example is constructed such that SC would be reached with a single DE if $\beta$ was unity. The used numbers are ($\mu$,$2\kappa$,$\gamma_{\rm{a}}$)=(\SI{5}{D},~\SI{100}{meV},~\SI{75}{meV}). We assumed a $Q$ of 20 and a resonance at \SI{2}{eV}. Further we used a vacuum decay rate $\gamma_{\rm{0}}$ of 1/(\SI{5}{ns}). These numbers result in ($g_{\rm{0}}$,$V_{\rm{M}}$,$\Gamma_{\rm{M}}$)=(\SI{43.3}{meV},~\SI{46.5}{nm^3},~\SI{2.31e6}{}). Only antennas with ultra-small gaps like bowtie antennas can, if any, provide such small mode volumes \cite{Koenderink2017Single-PhotonNanoantennas,Marquier2017RevisitingResonators}. The lines in Fig.~\ref{gamma_over_g_vs_N} cross unity at the needed number of DE to achieve SC, i.e., the critical emitter number $N_{\rm{c}}=\gamma\kappa/2g_{\rm{0}}^2$ \cite{Vahala2003,Kimble1998StrongQED}. Reaching SC is obviously more difficult if the $\beta$ factor is far below unity. We can further conclude that the impact of a limited $\beta$ factor is more pronounced for systems that are intended to reach SC with few DEs only.

We see that a small mode volume alone is not necessarily an advantage with respect to SC in plasmonic nanoresonators. For small $\beta$ factors, the only way to reach SC is to enhance the density of DEs in the active region of the resonator. However, there are constraints regarding the emitter density in real materials, for example when the distance between DEs approaches the F\"orster radius.
Therefore, a small mode volume may be even disadvantageous. This general consideration may loose its validity when $g$ or $\gamma_{\rm{\beta}}$ approach $\omega_{\rm{res}}$, i.e., when the Wigner-Weisskopf approximation is not valid anymore. Interestingly, a very similar consideration puts constraints on spasing using plasmonic nanoresonators \cite{Kewes2017LimitationsSpasers}.

With our proposed heuristic model introduced in the following section we can derive the minimum number of DEs to reach SC. From this the attenuation/gain constant $\alpha$ of a corresponding medium can be derived (e.g., in cm$^{-1}$). This allows for judging experimental feasibility by comparison with real material parameters reported in literature.

\section*{A Heuristic Model}
In this section, we introduce a heuristic modeling approach to numerically calculate relevant parameters of strongly coupled plasmonic systems. Our heuristic model is based on techniques and should be applicable to most Maxwell solvers. The heuristic model consists of two main steps. First, we compute effective rates using the framework of QNMs. QNMs allow for a direct calculation of the characteristic (effective) rates $g$, $\gamma$ and $\kappa$ (or effective complex mode volumes) only based on the geometry and the permittivities of the resonator and its surrounding. At this stage, it is already possible to discriminate between weak and strong coupling using the typical relations. In a second step one may use the derived numbers as input for a scattering simulation where a permittivity representation (short $\epsilon_{\rm{DE}}$-repr.) is used to model the emitters. This method has been applied frequently in the community \cite{Antosiewicz2014Plasmon-ExcitonCoupling} and describes the DEs by a frequency dependent permittivity typically in form of a Lorentzian. It allows for the calculation of scattering and absorption spectra including complicated experimental illumination and light detection situations. Please note, that the effective rates could also be used as input parameters in any appropriate analytic model to compute spectra, especially like discussed in Ref.~\cite{Lalanne2018LightResonances} where complex mode volumes are considered explicitly. 
Our heuristic model replaces a description of individual DEs by an effective medium. Here, all steps are executed with a frequency domain Maxwell solver (JCMsuite) based on finite elements. In averaging procedures, we make use of symmetries which exist in the specific examples we discuss below.

\subsection*{Computing Effective Rates} \label{sec:heuristic}
In situations where only few individual DEs are present with known positions and orientations, decay and coupling rates can be calculated explicitly (see example of nano particle on mirror below). In many experimentally relevant situations, however, e.g., for plasmonic particles that are coated with a layer of host material containing DEs (example of rod below), hundreds of DEs interact with the resonator, each of them in a different way. We aim to reduce the complexity of this problem by averaging over spatial position and orientation of numerous DEs (indicated by an overline, e.g., $\bar{\Gamma}_{\rm{M}}$), i.e., by calculating effective rates. For the example of the rod below, we assume random orientations of the DEs and a homogeneous spatial distribution. 

The framework of QNMs \cite{Sauvan2013,Kristensen2014} \cite{Lalanne2018LightResonances} allows for the calculation of parameters such as the mode volume $V_{\rm{M}}$, the modal Purcell factor $\Gamma_{\rm{M}}$ or the modal LDOS $\rho_{\rm{M}}$. Specifically, here we employed the Riesz projection method to derive the quantities \cite{Zschiedrich2018RieszNanoresonators}. This method allows to compute $\Gamma_{\rm{M}}$ fast and rigorously. In contrast to $\Gamma_{\rm{M}}$ and $\rho_{\rm{M}}$, the computation of the total Purcell factor $\Gamma_{\rm{tot}}$ and the total LDOS $\rho_{\rm{tot}}$ can be numerically very expensive since dipole sources need to be placed everywhere in the computational domain to compute the total emitted power for all relevant positions, orientations and frequencies. To circumvent this, we introduce the following heuristic averaging procedure. We start with the computation of the complex eigenfrequency and the Purcell factors $\Gamma_{\rm{M}}$ via the Riesz projection method. We determine the eigenfrequencies with JCMsuite which takes the dispersion relation into account. With the Riesz projection method we explicitly evaluate $\Gamma_{\rm{M}}$ for a DE in resonance with the fundamental plasmonic mode, i.e.,~$\omega_{\rm{0}}=\omega_{\rm{res}}$ at a single specific position $\textbf{r}_0$ and with a specific orientation only, using
\begin{equation}
\Gamma_{\rm{M}}(\textbf{r}_0)=-\frac{1}{2}\text{Re}(\textbf{E}(\textbf{r}_0)\cdot \bm{\mu})/\Gamma_0,
\end{equation}
where $\Gamma_0$ denotes the total decay rate in the corresponding homogeneous background medium. 
From this, we construct a spatial map of the modal Purcell factor using the relation $\Gamma_{\rm{M}}(\textbf{r})=\text{Re}(\eta(\textbf{r}) )\cdot \Gamma_{\rm{M}}(\textbf{r}_0)$. To this end we take the field computed with the eigenmode solver (which is not normalized \textit{a priori}) to define $\eta$ via
\begin{equation}
\eta(\textbf{r})=\frac{\Gamma_{\rm{M}}(\textbf{r})}{\Gamma_{\rm{M}}(\textbf{r}_0)}=\frac{\left(\textbf{E}(\textbf{r})\cdot\bm{\mu}\right)^2}{\left(\textbf{E}(\textbf{r}_0)\cdot\bm{\mu}\right)^2}.
\end{equation}
If the DEs are distributed, e.g., in a coating around the plasmonic resonator, we can use $\Gamma_{\rm{M}}(\textbf{r})$ to perform a spatial average to determine $\bar{\Gamma}_{\rm{M}}(\omega_{\rm{0}})$ right away. 

Next, to estimate $\bar{\Gamma}_{\rm{tot}}(\omega_{\rm{0}})$, we use the relation $\bar{\Gamma}_{\rm{tot}}(\omega_{\rm{0}})=\bar{\Gamma}_{\rm{M}}(\omega_{\rm{0}})/\bar{\beta}$ where $\bar{\beta}$ is left to be calculated. 
In order to do this, we compute $\Gamma_{\rm{M}}(\textbf{r}_i,\omega_{\rm{0}})$ and $\Gamma_{\rm{tot}}(\textbf{r}_i,\omega_{\rm{0}})$ for a small set of positions $\textbf{r}_i$. In this way, we save numerical resources, but at the prize of having to choose specific $\textbf{r}_i$. For the geometry of the examples which we study below, the DEs are randomly distributed in a coating layer around a plasmonic nanoresonator with cylindrical symmetry. In this situation, it is reasonable to choose the $\textbf{r}_i$ on a line perpendicular to the resonator's surface. We then derive the spatially averaged $\beta$ factor as $\bar{\beta}=\sum_{i}\frac{V_i}{V}\frac{\Gamma_{\rm{M}}(\textbf{r}_i,\omega_{\rm{0}})}{\Gamma_{\rm{tot}}(\textbf{r}_i,\omega_{\rm{0}})}$ where $V_i$ are volumes of the corresponding sublayers of the coating with thickness $|\textbf{r}_{i+1}-\textbf{r}_i|$ and $V$ is the volume of the complete coating. This estimation is optimistic if the $\textbf{r}_i$ are further chosen along a cylinder symmetry axis through a hotspot of the plasmonic mode, i.e., the regions of highest $\bar{\beta}$. 
So, we derive the effective Purcell factors $\bar{\Gamma}_{\rm{M}}(\omega_{\rm{0}})$ and $\bar{\Gamma}_{\rm{tot}}(\omega_{\rm{0}})$, respectively. From them we can calculate all the relevant decay rates for homogeneously broadened DEs.

At this point, we would like to point out that an inhomogeneously broadened DE or an emitter with phonon side bands can be represented by incoherent sums of homogeneously broadened emission lines. In this case, an additional spectral averaging would be required to derive effective rates instead of the evaluation applied here, that is based on $\Gamma$ at $\omega_{\rm{0}}$. Here, we refrain from considering such a more detailed modeling of the photophysics, since it would require to specialize on a certain emitter. Assuming merely homogeneously broadened emitters is, however, a best case scenario with respect to achieving SC.   

Finally, the quality factor $Q$ and the photon (or plasmon) decay rate $\kappa$ can be directly deduced from the complex eigenfrequency $\omega_{\rm{res}}=\omega_{\rm{res}}'+i\omega_{\rm{res}}''$ via $\kappa=-\omega_{\rm{res}}''$ and $Q=\omega_{\rm{res}}'/2\kappa$. Note, with the emitter number $N$ and the hosting volume of the DEs we can derive the density of DEs $\rho$ which allows to estimate the effective absorption or gain factor $\alpha$ of the densely packed DEs with the help of the absorption (emission) cross-section $\sigma$ of individual DEs: $\alpha=\rho\sigma$. 

\subsection*{Computing Spectra with the $\epsilon_{\rm{DE}}$-repr.} \label{sec:permittivity}

The second step is to perform extinction simulations using the aforementioned $\epsilon_{\rm{DE}}$-repr., where the DE is described by a frequency dependent permittivity in form of a Lorentzian. With this it is straight forward to calculate scattering, extinction, and absorption spectra with a Maxwell solver. In particular, it is possible to study the contribution of the plasmonic resonator and the DEs to the absorption separately, which is impossible in an experiment. The $\epsilon_{\rm{DE}}$-repr.~has been applied also to SC with nanostructures recently see, e.g., Ref.~\cite{Antosiewicz2014Plasmon-ExcitonCoupling}. The Lorentzian permittivity reads: 
\begin{equation}
\epsilon_{\rm{DE}} = \epsilon_{\rm{\infty}}+\frac{f \omega_{\rm{0}}^2}{\omega_{\rm{0}}^2-\omega^2-i\gamma_{\rm{a}} \omega}.
\label{permittivity}
\end{equation}
Here, $\epsilon_{\rm{\infty}}$ is a constant background permittivity, $\omega_{\rm{0}}$ is the resonance frequency, and $\gamma_{\rm{a}}$ represents the linewidth of the DEs. The parameter $f$ is crucial, as it controls the effective interaction strength with the DEs; $f$ not only represents the oscillator strength, but also relates to the density of DEs N/V in the simulation and must thus be calibrated accordingly. We connect $f$ to the number $N$ of DEs using parameters known from outside of the plasmonic mode's hotspot. According to Ref.~\cite{Torma2015StrongReview} this can be done when the dipole moment $\mu$ is known:
\begin{equation}
f=\frac{N}{V}\mu^2\frac{2}{3\epsilon_{\rm{0}}\hbar\omega_{\rm{0}}}.
\label{f}
\end{equation}

Further, as pointed out above care has to be taken to account for additional line broadening when the DE is brought in close proximity to a plasmonic resonator, i.e., an additional contribution $\gamma_{\rm{\beta}}$ has to be added to $\gamma_{\rm{a}}$ (in Eq.~\ref{permittivity}) to derive the total incoherent rate $\gamma=\gamma_{\rm{a}}+\gamma_{\rm{\beta}}$.

\section*{Strong Coupling in Plasmonic Resonators: Two Examples} \label{sec:examples}
Most of the resonant plasmonic structures which have been studied experimentally or theoretically show at least a cylindrical symmetry. For this reason, we focus on this symmetry as well. Further, cylindrical coordinates are implemented in the solver which reduces the computation time significantly. In particular, we investigate two structures: First, a nanoparticle on mirror (NPoM) structure as published recently \cite{Chikkaraddy2016Single-moleculeNanocavities}, where the authors report SC on the single DE level, and second a coated gold nanorod, which are frequently used in plasmonics to shift the resonance to the red side of the spectrum and to boost the $Q$ factor \cite{Dulkeith2004}. In all following examples, Purcell factors $\Gamma_{\rm{tot}}$, $\Gamma_{\rm{M}}$ and $\beta$ factors are calculated by our effective rates method that enter the computation of the spectra with the $\epsilon_{\rm{DE}}$-repr..

\subsection*{Nano Particle on Mirror (NPoM) of Ref.\cite{Chikkaraddy2016Single-moleculeNanocavities}} \label{sec:npom}

Chikkaraddy \textit{et al.} (Ref.~\cite{Chikkaraddy2016Single-moleculeNanocavities}) reported recently on a well defined experiment, where individual DEs (Methylene molecules) were dispersed sparsely on a flat gold film, with their dipole moments oriented perpendicular to the plane. Nanometre-sized gold spheres were then drop-casted, sometimes yielding a configuration where DEs were perfectly positioned within the tiny gap (width \SI{0.9}{nm}) between gold film and sphere. The DE's dipole moment ($\mu=\SI{3.8}{D}$) was then aligned along the dominating field vectors of a dipolar mirror-image mode providing an ultra-small $V_{\rm{M}}$.  

\begin{figure}
\includegraphics[width=0.99\columnwidth]{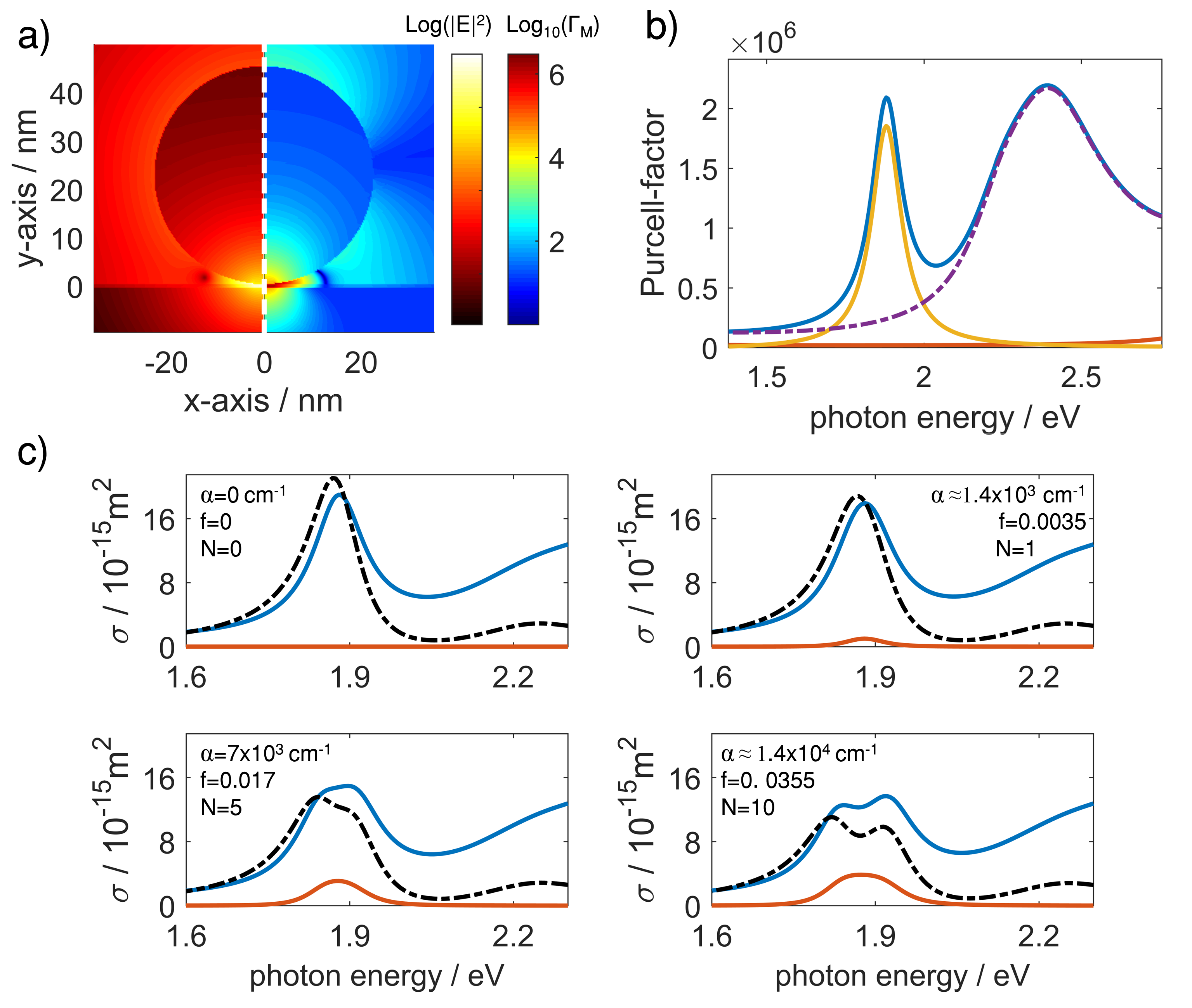}
\caption{
\label{fig_npom}NPoM with a sphere of radius $R=\SI{22.5}{nm}$. a) Log. intensity (left) of the field computed with the eigensolver, i.e., the resonant mirror-image mode and modal Purcell factor $\Gamma_{\rm{M}}$ for a DE in resonance oriented along the y-axis (right). The white dashed line marks the cylinder symmetry axis of the structure. b) 
Total Purcell factor $\Gamma_{\rm{tot}}$ (uppermost blue curve) and modal Purcell factor $\Gamma_{\rm{M}}$ (solid orange curve). 
In the difference curve between $\Gamma_{\rm{tot}}$ and $\Gamma_{\rm{M}}$ (dash-dotted purple), only the broad feature at around \SI{530}{nm} ($\approx$\SI{2.4}{eV}) remains, which corresponds to a pseudo-mode formed by a multitude of higher order modes (see text). The flat red curve shows for comparison the decay contribution of the two planar gold surfaces (sphere's surface and film). 
c) Calculated extinction (uppermost blue curve), scattering (black dashed), and absorption (red) spectra of DEs based on the $\epsilon_{\rm{DE}}$-repr.. For the absorption spectrum, only the absorption of the DEs is plotted. Note, the scattering cross section was multiplied by a factor of 7 for better visibility. Numbers in the inset refer to the gain, number $N$ of molecules, and corresponding $f$ parameter, respectively.
}
\end{figure}

Figure~\ref{fig_npom} summarizes our numerical results of this scenario. In Fig.~\ref{fig_npom}a) the field distribution found by the eigensolver, i.e., the mirror-image mode, with a $Q$ factor of 16.4 at \SI{665}{nm} ($2\kappa=\SI{118}{meV}$) and the corresponding modal Purcell factor $\Gamma_{\rm{M}}$ are shown. A hot spot in the gap can be clearly identified. In Fig.~\ref{fig_npom}b), we analyzed the DE's decay rate into different channels. The solid blue curve shows the energy dependence of the total Purcell factor, $\Gamma_{\rm{tot}}$. The pronounced resonant feature at \SI{665}{nm} ($\approx$\SI{1.8}{eV}) stems from the mirror-image mode strongly confined in the gap. This can be proven by plotting the modal Purcell factor $\Gamma_{\rm{M}}$ (solid orange curve) determined by a single resonant mode determined by the Riesz projection for a single eigenfrequency. The difference of the two curves (dashed-dotted purple curve) eliminates any resonant features at \SI{665}{nm}. The remaining broad feature at around \SI{530}{nm} ($\approx$\SI{2.4}{eV}) is a pseudo-mode formed by a multitude of higher order modes, that typically "condensate" around a specific energy depending on the metal; for gold, this occurs around \SI{2.35}{eV}, i.e., where intraband transitions occur \cite{Dulkeith2004}, yielding a broad and fast channel for DE decay \cite{Kewes2017LimitationsSpasers,Delga2014QuantumQuenching}. It is this pseudo-mode which limits the $\beta$ factor in the system.
Finally, we plot for comparison the contribution to the decay that stems from the sole gold substrate $\Gamma_{\rm{q}}$ (red line). This channel is present also far from any other resonance. There is an analytic approximation for this contribution given for a dipole at a small distance $r$ away from a planar metallic surface. In this case, there are two surfaces, so we double the contribution. The approximation is also accurate for dipoles close to metallic nanoparticles \cite{Moroz2010,Faggiani2015QuenchingDevices} and reads:
\begin{equation}
\Gamma_{\rm{q}}(r,\omega) = \frac{\omega}{(8 \epsilon_{\rm{d}} r^3)} (|\mu_{\rm{s}}|^2+\frac{|\mu_{\rm{p}}|^2}{2})\textrm{Im}\left[\frac{\epsilon_{\rm{m}}(\omega)-\epsilon_{\rm{d}}}{\epsilon_{\rm{m}}(\omega)+\epsilon_{\rm{d}}}\right]   
\label{quench} 
\end{equation}
where $\omega$ denotes the angular frequency, $\mu_{\rm{s}}$ and $\mu_{\rm{p}}$ the projection of the dipole moment onto orthogonal and parallel direction with respect to the metal surface, respectively, $\epsilon_{\rm{m}}$ is the permittivity of the metal, and $\epsilon_{\rm{d}}$ that of the host. We find a negligible contribution of the sole gold substrate, confirming our statement that the pseudo-mode is mainly responsible for degrading the $\beta$ factor.

Having revealed the different contributions to the decay channels we now check if SC is possible assuming a $\gamma_0$ of 1/(\SI{5}{ns}).
We find that $\gamma_{\rm{a}}=\SI{85}{meV}$ \cite{Chikkaraddy2016Single-moleculeNanocavities} outside of the plasmon mode's hotspot increases to \SI{116}{meV} inside.
The calculated values of ($\Gamma_{\rm{tot}}$, $\Gamma_{\rm{M}}$, $\beta$, $V_{\rm{M}}$, $g_{\rm{0}}$)=(\SI{2.09e6}, \SI{1.85e6}, \SI{88.6}{\%}, \SI{70.4}{nm^3}, \SI{27.65}{meV}) 
result in a value for the ratio $2\gamma\kappa/(2g_0\sqrt{N})^2$ of $N_{\rm{c}}$= 4.34 and consequently for a single DE a regime clearly below SC.
Note, even with $\gamma_{\rm{\beta}}$=0, $N_{\rm{c}}$ reaches 3.18.

In the experiment by Chikkaraddy \textit{et al.} (Ref.~\cite{Chikkaraddy2016Single-moleculeNanocavities}) scattering spectra were measured. We calculated extinction, scattering and absorption spectra using the $\epsilon_{\rm{DE}}$-repr.. We use a plane wave as light source, impinging from an angle of \SI{55}{\degree} from the top to mimic the dark field illumination described in Ref.~\cite{Chikkaraddy2016Single-moleculeNanocavities}. The DE is represented by a disk with \SI{4}{nm} diameter and \SI{0.9}{nm} height (the width of the gap) positioned in the center of the gap. The results are shown in Fig.~\ref{fig_npom}c). From the spectra we find that around 5-10 DEs are needed to enter the SC regime. This agrees very well with the calculated rates. One should note that such an estimate assumes that all DEs would fit into the antenna gap which has a physical size of only a few \SI{}{nm^3} (for example the \SI{4}{nm} disk and 5 DEs yield a density of $\rho=\SI{0.44}{nm^{-3}}$). 
This corresponds to an extreme dense packing where F\"orster processes are strong or the photophysics of the DEs might change completely. Assuming that the emission cross-section of Methylene blue ($\sigma\approx\SI{1.5e-20}{m^2}$) \cite{Milosevic2013OrientationSurface} does not change at such dense packing, the corresponding gain factor would reach $\alpha=\sigma \rho=\SI{66000}{cm^{-1}}$. 
Such extremely unrealistic values might explain why no splitting could be observed in the PL signal in the study of Chikkaraddy \textit{et al.} (Ref.~\cite{Chikkaraddy2016Single-moleculeNanocavities}). Further, our calculated scattering spectra with 5 or more DEs correspond very nicely to the experimental results presented by Chikkaraddy \textit{et al.} (Ref.~\cite{Chikkaraddy2016Single-moleculeNanocavities}). From this we conclude, however, that SC with single molecules was not observed. 
It is also important to note that a splitting in the scattering and extinction spectra may be observed (see Fig.~\ref{fig_npom}c)) before the critical number of DEs for SC is reached. A more solid prove of SC when observing spectra is to scrutinize the absorption spectra of the DEs \cite{Antosiewicz2014Plasmon-ExcitonCoupling}. Unfortunately, this is experimentally difficult.  
      
Nevertheless, there are two options to reach SC with a single emitter. Individual J-Aggregates (with $\mu \approx \SI{20}{D}$) may provide a sufficient oscillator strength. If it physically fits into the tiny gap of the NPoM resonator without being degraded, then SC may be observed. An alternative way is to eliminate the broad pseudo-mode, which would lead to even higher $\beta$ factors. This should be possible when using silver instead of gold. However, the plasmonic resonance would shift towards the blue side of the spectrum making it again harder to find a good DE with a large oscillator strength.

\subsection*{Gold Nanorod} \label{sec:rod}
\begin{figure}
\includegraphics[width=0.99\columnwidth]{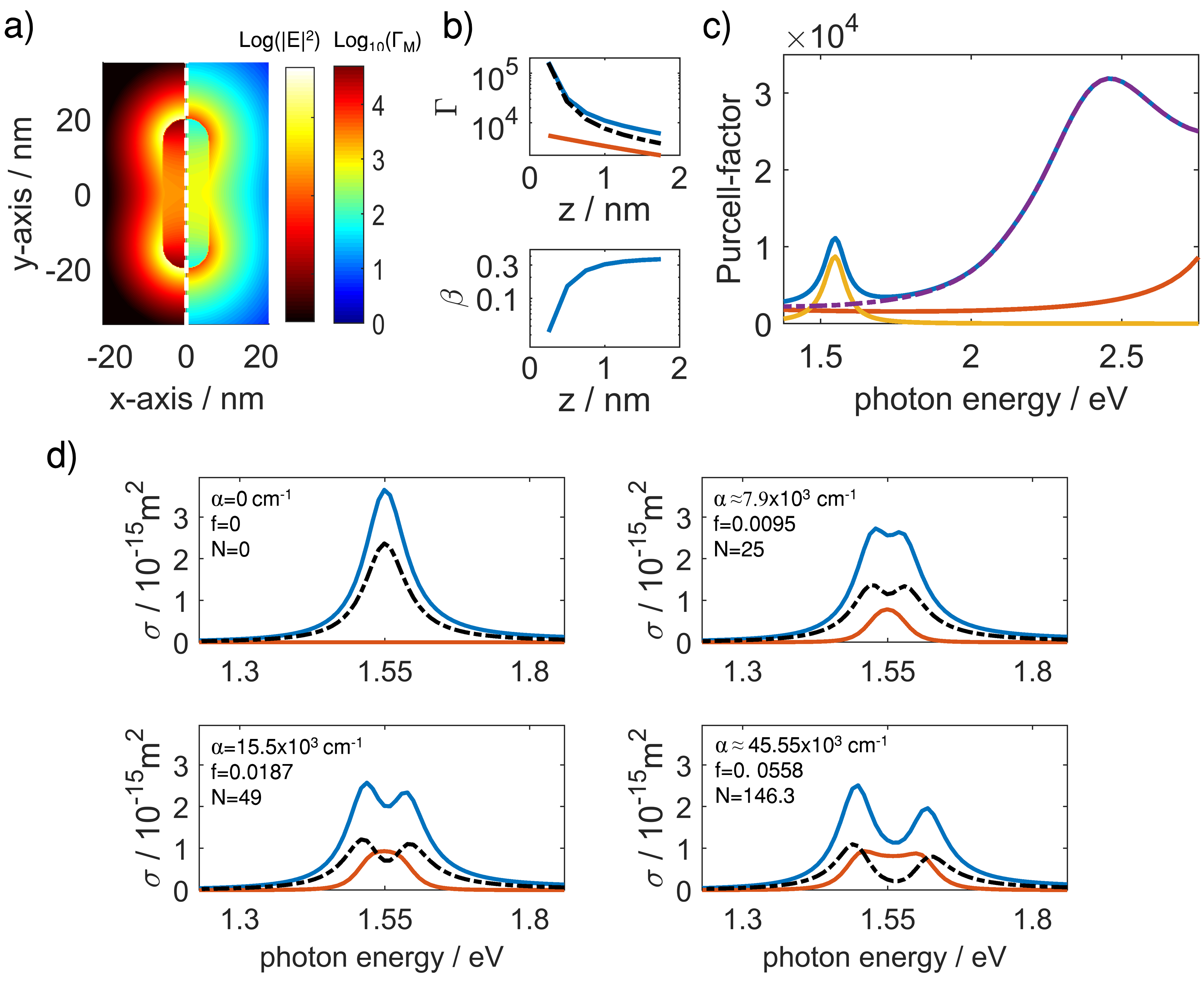}
\caption{
\label{fig_rod} Gold rod with a radius $R=\SI{6}{nm}$ and a length $L=\SI{40}{nm}$. 
a) Log. intensity (left) of the field computed with the eigensolver, i.e., the resonant mode and $\Gamma_{\rm{M}}$ for a DE in resonance (right). The white dashed line marks the cylinder symmetry axis of the rod. 
b) $\Gamma_{\rm{tot}}$ (blue), $\Gamma_{\rm{M}}$ (red) and their difference (black dashed) (upper plot) and ratio $\Gamma_{\rm{M}}/\Gamma_{\rm{tot}}$ (lower plot) as a function of the distance $z$ between the rod's surface and the DE along the symmetry axis.
c) Total and modal Purcell factors $\Gamma_{\rm{tot}}$  and $\Gamma_{\rm{M}}$, respectively, versus photon energy for a DE in the hotspot of the plasmonic mode (\SI{1}{nm} above the rod on the sym.-axis). A narrow and a broad peak are found for $\Gamma_{\rm{tot}}$ (uppermost blue curve), that stem from the resonant mode (orange) and a pseudo-mode at around \SI{530}{nm} (dash-dotted purple). The flat red curve shows for comparison the contribution of DE decay due to a single planar gold film. The inset shows a zoom to the spectral region of the fundamental plasmonic resonance. 
d) Calculated extinction (uppermost blue curve), scattering (black dashed), and absorption (red) based on the $\epsilon_{\rm{DE}}$-repr.. 
For the absorption spectrum, only the absorption of the DEs is plotted. Numbers in the inset refer to the gain, number $N$ of molecules, and corresponding $f$ parameter, respectively.
}
\end{figure}

The second nanostructure we investigate is a plasmonic nanorod. Its one-dimensional geometry provides the possibility to match two modes with vastly different resonance frequencies almost independently by changing the rod's width and length. At the same time a rod resonator already resembles a waveguide, which facilitates out-coupling of plasmonic excitations to on-chip structures. 
We model a gold rod with a diameter of \SI{6}{nm} and a length of \SI{40}{nm} with rounded end facets surrounded by a medium with a refractive index of $n=1.5$. A coating of the rod with DEs (e.g., organic molecules) of \SI{2}{nm} thickness is assumed. For this example, we choose DEs with typical values as found for J-Aggregates with a width of $\gamma_{\rm{a}}=\SI{50}{meV}$,$\epsilon_{\infty}=1.5^2$, and a dipole moment of $\mu=\SI{20}{D}$.
Figure \ref{fig_rod} summarizes our calculations for this setup. First, we found a fundamental plasmonic resonance at \SI{800}{nm} with a $Q$ factor of $18$. This $Q$ factor is as high as one can get with such small gold particles as predicted by the quasi static description \cite{Wang2006}. Its intensity distribution shows a dipolar behavior and is plotted in Fig.~\ref{fig_rod}a). The chosen parameter of the rod were motivated by spectrally separating the dipole mode and the pseudo-mode, which limited the performance of the NPoM structure discussed in the previous section. Indeed, the pseudo-mode is found far away at around \SI{530}{nm}.

In Fig.~\ref{fig_rod}b) and c) we analyzed the DE's decay rate into different channels using our effective rates model. All calculated Purcell factors ($\Gamma_{\rm{tot}}$ and $\Gamma_{\rm{M}}$) include averaging over the DE's orientation. In the upper plot of Fig.~\ref{fig_rod}b), we show the dependency of the total and modal Purcell factor $\Gamma_{\rm{tot}}$ (blue solid line) and $\Gamma_{\rm{M}}$ (red solid line), respectively, as a function of distance $z$ between the rod surface and the DE along the symmetry axis.  The lower plot shows the $\beta$ factor, $\beta=\Gamma_{\rm{M}}/\Gamma_{\rm{tot}}$. 
It is apparent that the $\beta$ factor quickly drops far below unity when approaching the rod's surface. This is since coupling to higher order modes (or the pseudo mode) increases much faster than the coupling to a single resonant mode when approaching the particle surface \cite{Kewes2017LimitationsSpasers,Moroz2010}. 
The energy dependency of $\Gamma_{\rm{tot}}$  and $\Gamma_{\rm{M}}$  for a DE in the hotspot of the plasmonic mode are shown in Fig.~\ref{fig_rod}c).  
Similar as in the NPoM structure a narrow and a broad peak are found. The first one (at \SI{800}{nm}) stems basically from one resonant mode whereas the second one is again the pseudo-mode (at \SI{530}{nm}).
The flat red curve in Fig.~\ref{fig_npom}c) shows for comparison the contribution of DE decay due to a single planar gold film in contrast to \ref{sec:npom} (showing the contribution of two films), where the DE was positioned in a gap.
We already pointed out that due to the rod's geometry, both modes are split further apart as compared to the case of the NPoM structure. However, decay to unwanted modes still plays an important role. This is, because the modal Purcell factor is significantly weaker as for the NPoM resonator with its gap. The contribution of a single gold interface is in this example as high as that of the pseudo-mode; both channels drastically limit the $\beta$ factor. 

After revealing the different contributions to the decay channels we now consider the possibility of SC in the nanorod resonator with DE coating.
We find the following effective values of ($\bar{\Gamma}_{\rm{tot}}$, $\bar{\Gamma}_{\rm{M}}$, $\bar{\beta}$, $\bar{V}_{\rm{M}}$, $\bar{g}_{\rm{0}}$)=(\SI{6700}, \SI{1794}, \SI{26.8}{\%}, \SI{1.16e5}{nm^3}, \SI{3.04}{meV}). Taking these parameters to evaluate the required critical number $N_{\rm{c}}$ of DE to reach the SC regime results in a quite large value of $N_{\rm{c}}=146.3$ corresponding to a density of $\rho$=\SI{0.045}{nm^{-3}}. The broadening due to additional decay channels reaches ($\gamma_{\rm{\beta}}=\SI{12.3}{meV}$).

Finally, we calculated extinction, scattering and absorption spectra using our $\epsilon_{\rm{DE}}$-repr.. The light sources used in the Maxwell solver correspond to an extinction measurement in solution, i.e., the extinction spectra are averaged for the different possible particle orientations with respect to the plane wave Poynting vector and polarisation.
The results are shown in Fig.~\ref{fig_npom}d) for different numbers of molecules in the rod's coating layers. We find again the feature that a splitting in the extinction and scattering spectra appears long before the SC regime is reached. If we take the onset of splitting in the absorption spectrum of the DEs as a more solid signature of SC, we see that about 140 DEs are needed. This again agrees nicely with the calculation of the critical number of DEs of  $N_{\rm{c}}=146.3$ above. This number of DE relates to a gain factor of $\alpha$=\SI{8.85e5}{cm^{-1}} using a $\sigma=\SI{2.15e-14}{cm^2}$ for TDBC aggregates \cite{Zengin2015RealizingConditions}. 

\section*{Discussion}
In conclusion, we have discussed potential sources of additional broadening of DEs in SC with plasmonic cavities.
The broadening originates from fast relaxation rates near metals, that are typically not present in photonic systems for SC. To include this broadening in widely available simulation tools like typical Maxwell solvers, we proposed a way of computing effective coupling rates $\bar{g}$ and with it a modified (enhanced) damping rate $\bar{\gamma}$.  Our heuristic methods are applicable even to complicated structures that are hard to model otherwise. The effective rate  $\bar{\gamma}$ might be compared to the effective coupling $\bar{g}$ and the resonator damping rate $\kappa$ to evaluate whether the conditions of SC are fulfilled. The derived effective rates $\bar{\gamma}$ can also be used to describe the damping of a Lorentzian permittivity in simulations of extinction, scattering or absorption spectra. Further, the required number of DEs or the density of DEs can be deduced. In this way, corresponding gain factors can be calculated and compared to values of real materials. We conclude that most systems for SC based on nano particles would require unrealistically high gain factors to yield an observable splitting in PL studies. We pointed out that for the design of efficient nanostructures for SC one must optimize $\beta$ and $g_0$ at the same time, especially when one strives for SC with a few DEs only. Our findings show that reports on SC on the single DE level are at least ambiguous at present. The fact, that our spectral simulations match the experimental findings of Ref.~\cite{Chikkaraddy2016Single-moleculeNanocavities} so nicely, indicates that our heuristic model is indeed an appropriate tool to study light-matter interaction even in such extreme situation with ultra-small mode volumes. Along with the transparency of the heuristic model comes the possibility to identify optimization routes like the ones discussed above. Considering the potential for optimization of NPoM or similar designs, we can thus conclude that it is in principle possible to reach the strong coupling limit with single emitters only. 

Generally, in order to reduce additional broadening in plasmonic particle-based resonators several approaches are possible. When using gold nanostructures one should select DEs that are far away from the pseudo-mode around \SI{530}{nm}. Further, it is possible to some extend to construct a plasmonic resonator that features a density of states $\rho_{\rm{tot}}$ that significantly exceeds the contribution from off-resonant channels. In principle, plasmon resonators with quasi unity $\beta$ factor are possible \cite{Zhang2017ANanoresonator}, though the physical volume available for hosting DEs will be very limited. Consequently, one has to find a compromise between usable volume and effective rates for each material system and purpose. Our conclusion, that both, $\beta$ factor and $g_{\rm{0}}$ must be optimized simultaneously is further in line with Ref.~\cite{Delga2014QuantumQuenching} where the authors observed an effective detuning of the hybridized states away from the pseudo-mode. The maximum of the $\beta$ factor will shift with respect to the peak of the modal LDOS when the resonator mode spectrally approaches a pseudo mode or the impact of the pseudo mode is strengthened when the emitters approach the metal surface. We can further conclude that rod-like or waveguide structures of finite length supporting Fabry-P\'erot resonances made from silver should be preferred over nanoparticles when aiming at SC with many DEs. The pseudo-mode of silver nanostructures is spectrally substantially shifted to the blue side of the spectrum compared to gold. Such nanowire-based Fabry-P\'erot resonators allow for tuning length and cross section and thus $Q$ factor and coupling strength $g_{\rm{0}}$ more freely. Especially $Q$ can be enhanced beyond the limits found for particles \cite{Wang2006}. 
Apart from fundamental studies of SC, real world applications may consider designs exploiting Babinet's principle. Channel plasmons as found in drilled holes in metal films could serve as superior Fabry-P\'erot-like plasmon resonators for modifications of photo physics or opto-electronics properties of DEs, organic molecules or polymer materials \cite{Hutchison2012ModifyingFields,Memmi2017StrongVibrations}.

\section*{Acknowledgement}
We acknowledge financial support of Einstein Foundation Berlin (ECMath, project OT9) and the  German Research Foundation (DFG, CRC 787, project B4 and CRC 951, project B2)




\providecommand{\latin}[1]{#1}
\makeatletter
\providecommand{\doi}
  {\begingroup\let\do\@makeother\dospecials
  \catcode`\{=1 \catcode`\}=2\doi@aux}
\providecommand{\doi@aux}[1]{\endgroup\texttt{#1}}
\makeatother
\providecommand*\mcitethebibliography{\thebibliography}
\csname @ifundefined\endcsname{endmcitethebibliography}
  {\let\endmcitethebibliography\endthebibliography}{}

\end{document}